\documentclass[reprint,
superscriptaddress,
 amsmath,amssymb,
 aps,
12pt,
onecolumn
]{revtex4-1}
\usepackage{setspace}
\usepackage{graphicx}
\usepackage{dcolumn}
\usepackage{bm}
\usepackage[colorlinks=true,linkcolor=blue]{hyperref}
\hypersetup{
    colorlinks=true,
    linkcolor=blue,
    citecolor=blue,
    filecolor=magenta,      
    urlcolor=blue,
}

\begin{document}

\preprint{APS/123-QED}

\title{Strange Nonchaotic Attractor in an Unforced Turbulent
Reactive Flow System }

\author{Thonti Beeraiah}
\author{Shruti Tandon}
\affiliation{%
  Department of Aerospace Engineering, Indian Institute of Technology Madras, Chennai 600 036, India}%
\affiliation{
  Centre of Excellence for Studying Critical Transition in Complex Systems, Indian Institute of Technology Madras, Chennai 600 036, India}
\author{Premraj Durairaj} 
\affiliation{Institute of Systems Science, Huaqiao University, Xiamen, 361021, China}%
\author{R. I. Sujith}
\affiliation{%
  Department of Aerospace Engineering, Indian Institute of Technology Madras, Chennai 600 036, India}%
\affiliation{
  Centre of Excellence for Studying Critical Transition in Complex Systems, Indian Institute of Technology Madras, Chennai 600 036, India}


\date{\today}

\begin{abstract}
We discover strange nonchaotic attractor (SNA) through experiments in an unforced system comprising turbulent reactive flow. While models suggest SNAs are common in dynamical systems, experimental observations are primarily limited to systems with external forcing. We observe SNA prior to the emergence of periodic oscillations from chaotic fluctuations. In complex systems, self-organization can lead to order, and inherent nonlinearity can induce chaos. The occurrence of SNA, which is nonchaotic yet nonperiodic in one such complex system, is intriguing.
\end{abstract}

\maketitle

\textit{Introduction}\textemdash
Strange nonchaotic attractor (SNA) is a non-trivial dynamical state that has attracted sustained interest in the nonlinear dynamics community  \cite{grebogi1984strange, ding1989evolution, feudel1995strange, romeiras1987strange, heagy1991dynamics, zhou1992observation}. SNAs exhibit intricate fractal geometries akin to those observed in chaotic systems. However, unlike chaotic systems, SNAs do not exhibit exponential sensitivity to initial conditions and are hence characterized as nonchaotic. Typically, a positive largest Lyapunov exponent indicates sensitivity to initial conditions; for SNAs, this exponent is nonpositive \citep{grebogi1984strange}.\\
\hspace*{5mm} We discover the state of SNA in a system comprising turbulent reactive flow. Our finding is particularly intriguing given that our system is not externally forced and is rather a self-organized complex system owing to a web of inter-subsystem interactions \citep{sujith2021thermoacoustic, tandon2023multilayer}. Turbulent flows are ubiquitous in nature and consist of eddies ranging from very small scale to very large scale \cite{tennekes1972first}. These eddies interact nonlinearly, making turbulent flow a quintessential complex system \cite{pierrehumbert2022fluid}. Introducing an exothermic reaction into a turbulent flow adds to the intricacy and multifaceted nature of the system.\\
 \hspace*{5mm} Turbulent reactive flows occur in natural systems, including stars, which are thermonuclear reactors composed of plasma, bounded by gravity. These stars undergo turbulent mixing \cite{arnett2016key}, and in rotating stars, this turbulent mixing generates a magnetic field which, in turn, establishes a nonlinear interaction between turbulent mixing, rotation, and magnetic field. Such stars may exhibit chaotic fluctuations \cite{PhysRevLett.74.842} or unstable pulsation modes \cite{arnett2016key}. Turbulent reactive flows also exist in human-made systems, such as combustors in gas turbine and rocket engines. Turbulent reactive flows in combustors comprise subsystems, namely the hydrodynamic field, heat release (flame) field, and acoustic field that interact in a nonlinear manner, increasing the overall complexity \citep{ChuKovasznay1958}. During the stable operation in such systems, the acoustic pressure fluctuations ($p'$) are low-amplitude and aperiodic in nature and are characterized as high-dimensional chaos \citep{nair2013loss,tony2015detecting}. During this state, the subsystems are desynchronized \citep{pawar2017thermoacoustic, mondal2017onset}. In contrast, during unstable operation, high-amplitude periodic acoustic pressure oscillations occur, characterized as limit cycle oscillations (order). Order emerges from the chaotic fluctuations when positive feedback is established between the heat release rate and the acoustic field, leading to their synchronization \cite{sujith2021thermoacoustic}. In these systems, the presence of chaotic and ordered states, and the transition from chaotic to ordered states via a state of intermittency\textemdash where high amplitude periodic epochs are interspersed with low amplitude aperiodic fluctuations\textemdash is well established  \cite{nair2014intermittency,10.1063/5.0001900}. However, the discovery of the state of SNA, which is nonchaotic yet nonperiodic, in such a complex system is surprising.\\
\hspace*{5mm}Typically, SNAs are found in dynamical systems driven by external quasiperiodic \cite{grebogi1984strange, ding1989evolution, feudel1995strange, romeiras1987strange, heagy1991dynamics, zhou1992observation, ding1994phase, bondeson1985quasiperiodically, heagy1994birth, prasad1997intermittency, wang2004strange, jalnine2007realization} or periodic forcing \cite{anishchenko1996strange, doi:10.1142/S0218127401002109}. SNAs have also been observed in experimental systems with periodic forcing  \cite{guan2018strange, Chithra2021}. Furthermore, a few studies using mathematical models have demonstrated that neither quasiperiodic force nor periodic force is essential for the occurrence of SNA \cite{negi2001plethora, kumarasamy2022strange}. However, to date, only two physical systems have been reported to exhibit SNAs without being subjected to external force: a pulsating star KIC $5520878$ network \cite{lindner2015strange} and a laminar thermoacoustic system \cite{premraj2020strange}.\\
\hspace*{5mm}In this letter, we present the experimental discovery of state of SNA in a highly complex turbulent reactive flow system. In the experiments, we increase the Reynolds number ($Re$) as a bifurcation parameter in a quasi-static manner and observe a transition from low-amplitude aperiodic fluctuations to high-amplitude periodic oscillations. We show that the state of SNA occurs prior to the onset of high-amplitude periodic oscillations. While the computation of Lyapunov exponents is generally useful for determining the nature of dynamical states, ranging from regular to chaotic, it is often unreliable for noise-contaminated experimental data \cite{PhysRevLett.65.1523}. Therefore, we employ Fast Fourier Transform (FFT) and phase space reconstruction to distinguish different dynamical states. We then apply the $0-1$ test \citep{gottwald2004new} and correlation dimension test \citep{grassberger1983characterization} to characterize these states. Finally, to confirm the presence of SNA, we use singular continuous spectrum analysis \citep{Pikovsky_1994}.\\  
\hspace*{10mm}\textit{Experimental setup}\textemdash
\begin{figure}
\includegraphics[scale = 0.35]{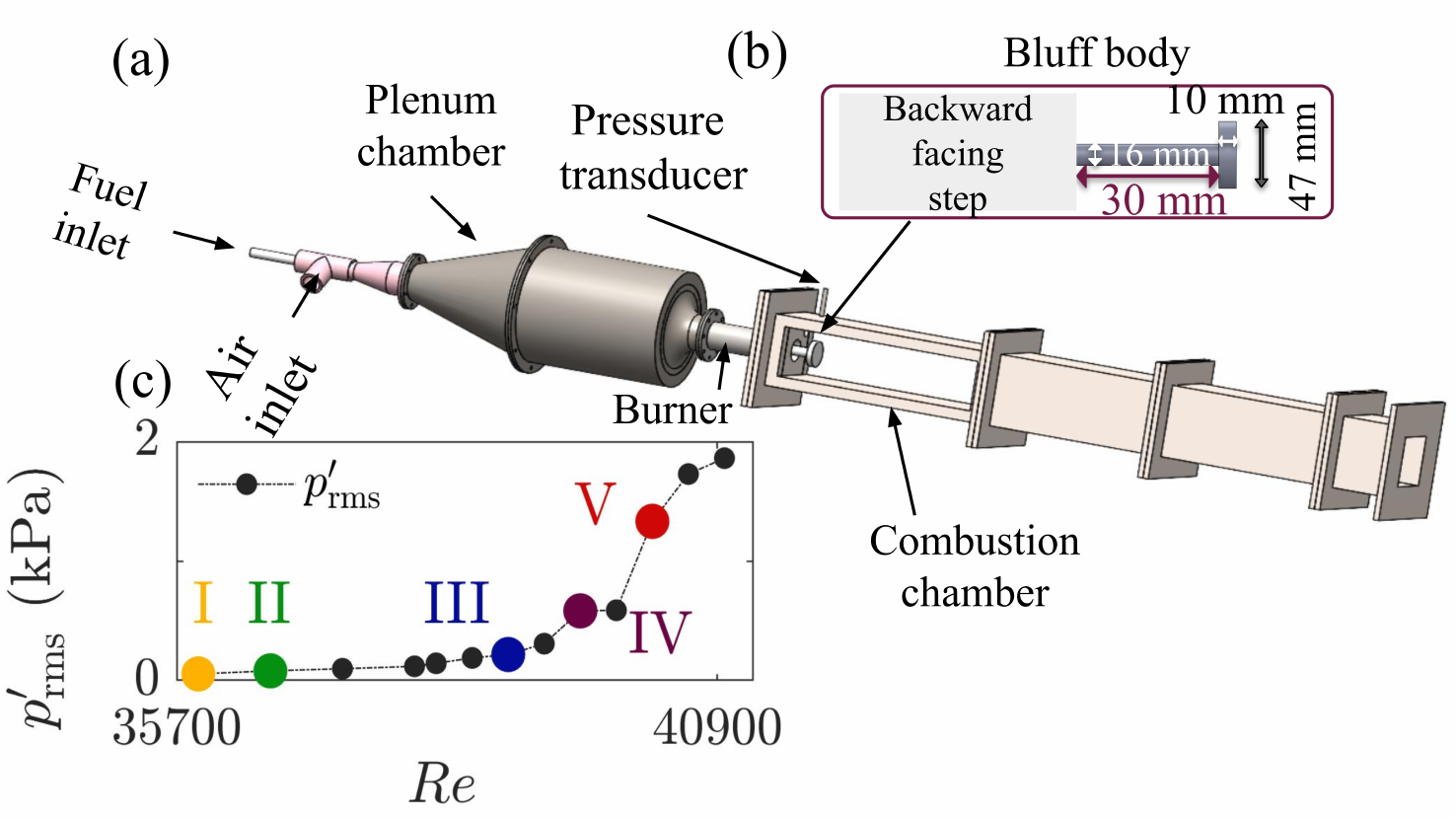}
\caption{\label{setup}(a) Schematic of the experimental setup, (b) bluff body mounted on a shaft fixed at $30$ mm from the backward-facing step, and (c) variation of root-mean-square of acoustic pressure fluctuations ($p'$) as a function of $Re$ during the transition, with the highlighted points indicating various dynamical states present.}
\end{figure}
We perform experiments in a lab-scale backward-facing step turbulent combustor, using a bluff body as a flame stabilizer, as depicted in figure~\ref{setup}\textcolor{blue}{a}. The bluff body is fixed at a location of $30$ mm from the backward-facing step throughout the transition. The setup comprises three main components: a plenum chamber, a burner, and a combustion chamber. The combustion chamber has a square cross-section measuring $90$ mm $\times$ $90$ mm and a length of $1100$ mm. The plenum chamber connects to the burner, which further leads to the combustion chamber. Gaseous fuel is injected into the air in the burner section, creating a partially premixed fuel-air mixture by the time it reaches the combustion chamber, where all reactions occur. To vary the Reynolds number ($Re$), we fix the fuel flow rate and increase the air flow rate using mass flow controllers (Alicat Scientific, MCR). The uncertainty in the measurement of the flow rate is $\pm$(0.8\% of reading + 0.2\% of full scale), resulting in a maximum uncertainty of $\pm2.5\%$ in $Re$. We flush-mount a piezoelectric pressure transducer (PCB$103$B$02$) on the combustion chamber at $40$ mm from the backward-facing step to measure acoustic pressure fluctuations ($p'$). The sensor has an uncertainty of $\pm 0.15$ Pa. During the transition, we measure the acoustic pressure oscillations ($p'$) at a sampling rate of $20$ kHz.\\
\textit{Results}\textemdash
\begin{figure*}
\includegraphics[scale = 0.6]{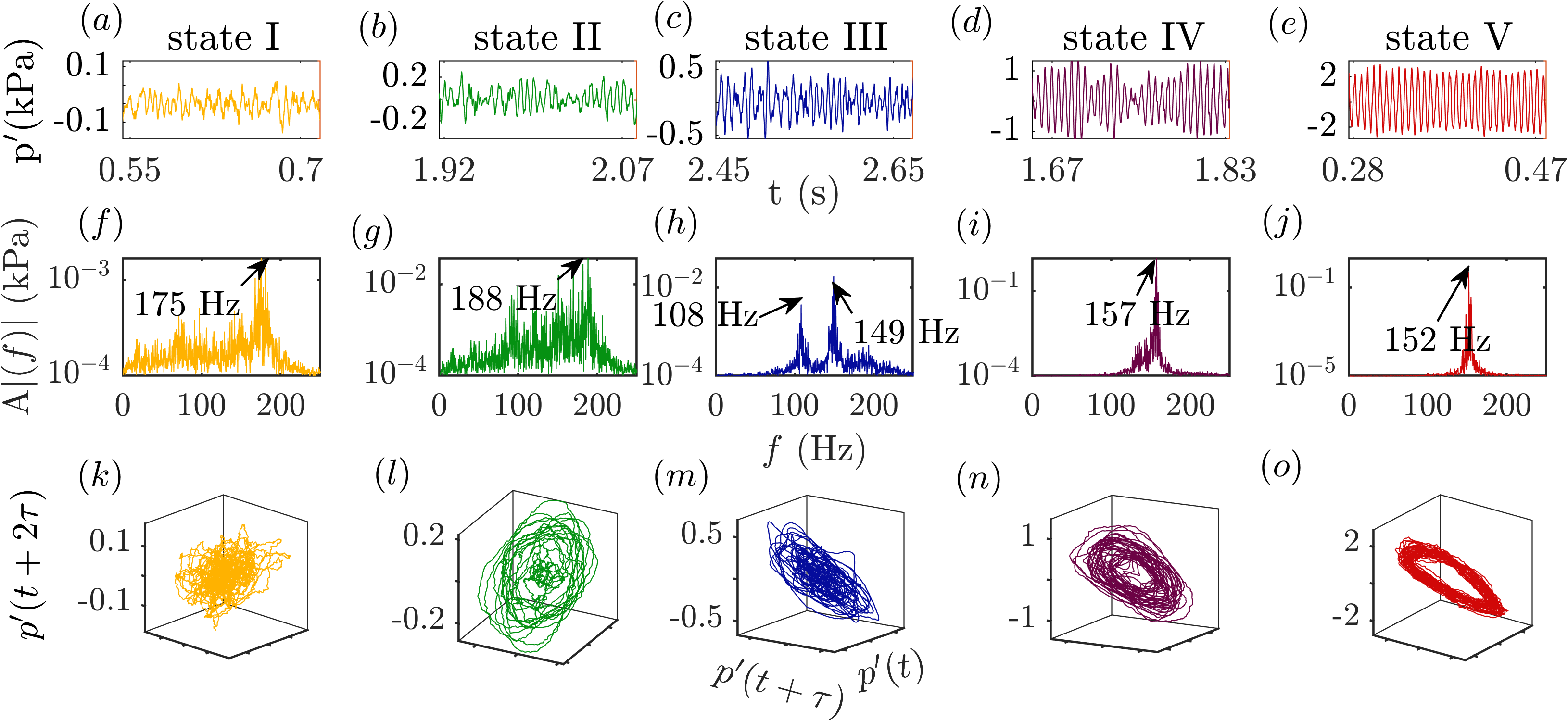}
\caption{\label{FFT_PS} Time series (a-e), Amplitude spectrum (f-j), and reconstructed phase space (k-o) of the five dynamical states. During the transition, periodicity emerges, as reflected in the amplitude spectrum and reconstructed phase space.}
\hspace*{10mm}\end{figure*}
In our experiments,  we vary $Re$ from $28,000$ to $44,000$ and observe a transition from low-amplitude aperiodic fluctuations to high-amplitude periodic oscillations (Fig.~\ref{setup}\textcolor{blue}{c}). We primarily use Fast Fourier Transform and phase space reconstruction of $p'$ to distinguish the dynamical states present, labeled as states I to V (Fig.~\ref{FFT_PS}). As the transition occurs from state I to V, we observe an increase in the amplitude of fluctuations in $p'$ (Fig.~\ref{FFT_PS}\textcolor{blue}{a-e}). Concurrently, the periodicity of the signal increases, culminating in the emergence of a single prominent peak in the amplitude spectrum during state V (Fig.~\ref{FFT_PS}\textcolor{blue}{j}) compared to the broad peak during state I (Fig.~\ref{FFT_PS}\textcolor{blue}{f}). To further characterize these states, we apply $0 - 1$ test and correlation dimension test. Finally, we employ the singular continuous spectrum test to confirm the SNA, which we identify through the preceding analyses.
\paragraph{$0-1$ test}
Originally, \citet{gottwald2004new} proposed this method to discriminate between regular and chaotic time series. The initial step of the $0-1$ test is to transform the original signal into a new space where the underlying dynamics becomes more apparent. For the same purpose we calculate the translation variables ($p_n(n)$ and $q_n(n)$), as defined by,
\begin{subequations}
    \begin{equation}
        p_n(n) = \sum_{i=1}^{n} p'(i) \cos(ic),
    \end{equation}
    \vspace{-0.5cm}
    \begin{equation}
        q_n(n) = \sum_{i=1}^{n} p'(i) \sin(ic), n = 1,2,3,...,
    \end{equation}
    \label{translation var}
\end{subequations}
where, $p'(i)$ denotes the acoustic pressure fluctuations at instant $i$ and $c$ is a parameter selected arbitrarily within $\pi/5$ to $4\pi/5$. The dynamics of variables $p_n(n)$ and $q_n(n)$ are closely related to the discrete variable $n$, which is significantly smaller than the length of the signal ($N$) (typically $n = N/10$). To quantify the evolution of the trajectory in the $[p_n(n), q_n(n)]$ plane as $n$ increases, the mean square displacement $D(n)$ is utilized, defined as:
\begin{equation}
D(n) =\frac{1}{N} (\sum_{i=1}^N [p_n(i+n) - p_n(i)]^2 + [q_n(i+n) - q_n(i)]^2).
\end{equation}
To address convergence issues with $D(n)$, a modified mean square displacement $M(n)$ \cite{gottwald2009implementation, gottwald2009validity} is proposed, which is calculated as:
\begin{equation}
    M(n) = D(n) - V_{osc}(c,n),
\end{equation}
where, $V_{osc}(c,n) = \sum_{i=1}^n p'(i) \frac{1-\cos(ic)}{1-\cos(c)}$, is an oscillatory term.  The asymptotic growth rate $K_c$ is determined through linear regression fit of $M(n)$ as:
\begin{equation}
    K_c = \lim_{n \to \infty} {\frac{\log M(n)}{\log n}}.
    \label{k}
\end{equation}
The dynamical states are distinguished using the variation of $M(n)$ and magnitude of $K_c$. The value of $M(n)$ grows linearly for chaotic dynamics owing to sensitivity to initial conditions while it remains bounded for periodic dynamics \cite{gottwald2009implementation, gottwald2009validity}. This behavior causes the value of $K_c$ to fall within the range of $1$ to $0$, where a value of $K_c = 1$ suggests chaotic dynamics and $K_c = 0$ implies periodic behavior. For the dynamical state of SNA, the value of $K_c$ is distributed between $0$ and $1$, reflecting its nature as a state that is neither chaotic nor periodic.\\ 
\begin{figure*}
\includegraphics[scale = 0.57]{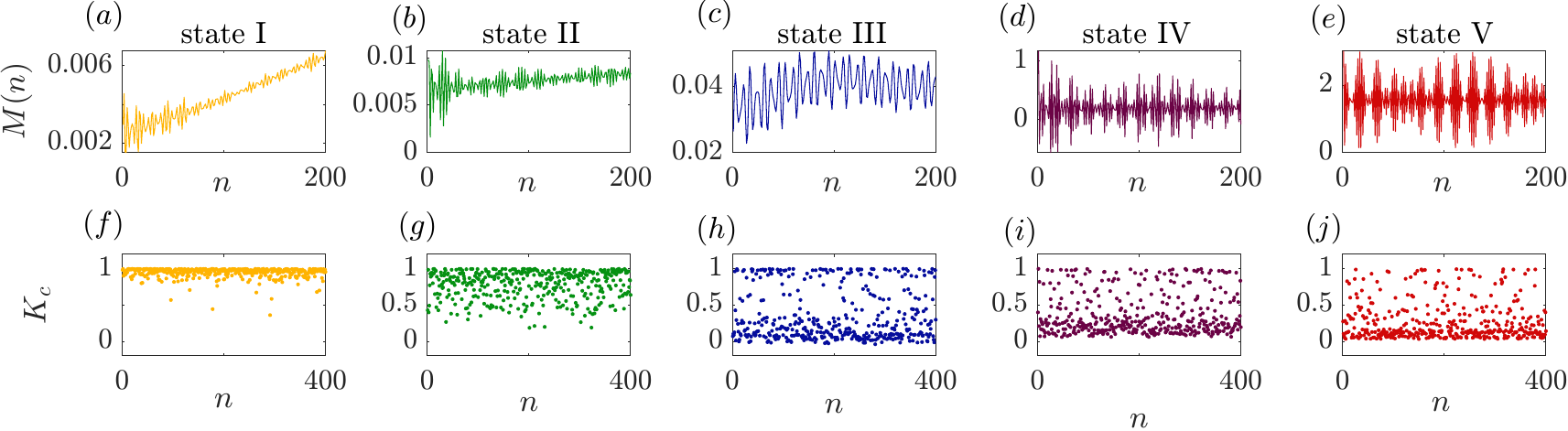}
\caption{\label{0-1} Modified mean square displacement ($M(n)$) (top row) and asymptotic growth rate ($K_c$) (bottom row) during the dynamical states of (a, f) I, (b, g) II, (c, h) III, (d, i) IV, and (e, j) V. The value of $K_c$ decreases gradually from $1$ to $0$ during the transition from state I to V, indicating the emergence of periodicity.}
\end{figure*}
\hspace*{10mm}In figure~\ref{0-1}, we plot $M(n)$ and $K_c$ for the dynamical states I-V. For state I, the value of $M(n)$ increases linearly on average(Fig.~\ref{0-1}\textcolor{blue}{a}), and the value of $K_c$ is fairly lying close to $1$ (Fig.~\ref{0-1}\textcolor{blue}{f}) depicting a chaotic nature of the state. During state II also, the value of $M(n)$ increases linearly on average with slight fluctuations (Fig.~\ref{0-1}\textcolor{blue}{b}) and the values of $K_c$ distributed just below $1$ (Fig.~\ref{0-1}\textcolor{blue}{g}), showing a chaotic nature.\\
\hspace*{10mm}For state III, the value of $M(n)$ either increases or decreases in a bounded range on average as evident in figure~\ref{0-1}\textcolor{blue}{c} and the value of $K_c$ is distributed between $1$ and $0$ (Fig.~\ref{0-1}\textcolor{blue}{h}), depicting that the state III as neither chaotic nor periodic. During state IV, the value of $M(n)$ becomes bounded and the average value remains constant with fluctuations present (Fig.~\ref{0-1}\textcolor{blue}{d}). The value of $K_c$ distributed just above $0$ (Fig.~\ref{0-1}\textcolor{blue}{i}), showing a periodic behavior. During state V also, the value of $M(n)$ bounded and the average value remains constant with fluctuations present (Fig.~\ref{0-1}\textcolor{blue}{e}), the values of $K_c$ are fairly lying close to $0$ (Fig.~\ref{0-1}\textcolor{blue}{j}), delineating a clear periodic behavior, compared to the previous state. 
\paragraph{Correlation Dimension}
\begin{figure*}
\includegraphics[scale=0.55]{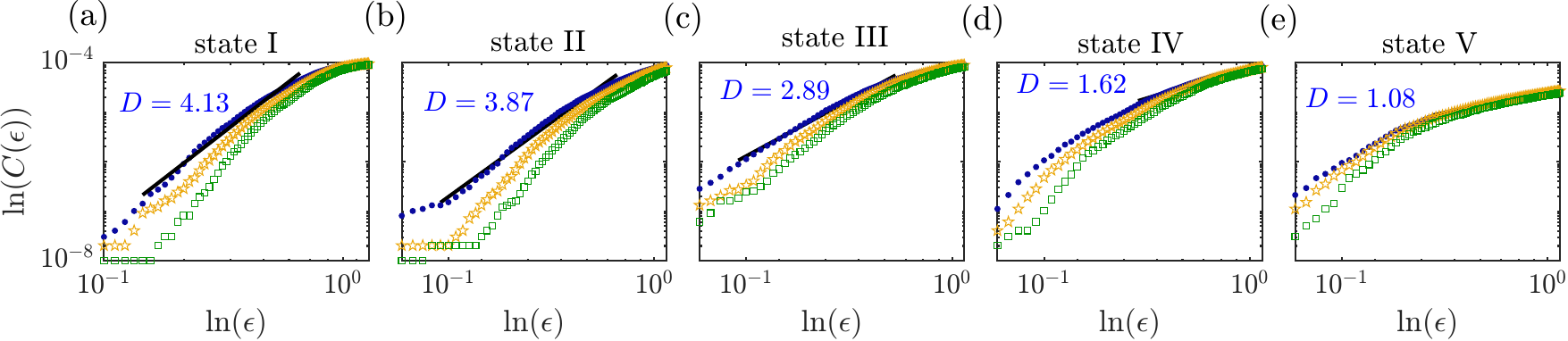}
\caption{\label{corr_sum} (a-e) Log-log plots showing the variation of correlation sum (C($\epsilon$)) with $\epsilon$ for the states I to V. The plots are shown for the embedding dimensions ranging from $5$(\textcolor{blue}{$\bullet$}), $6$ (\textcolor{yellow}{$\star$}) and $7$ (\textcolor{green}{$\Box$}). The solid line represents the power law fit for the straight-line portion of this plot, indicating the value of the correlation dimension ($D$). The value of $D$ decreases as the system transitions from state I to state V, indicating the decrease of complexity.}
\end{figure*}
The correlation dimension is a topological measure of the fractal dimension of an attractor \cite{book:kantz_schreiber_2003}. We compute the correlation dimension using the algorithm proposed by \citet{grassberger1983characterization}. The initial step is to calculate the delayed vectors $s_n = \{p_i'(m)\}_{i=1}^{N-(m-1)\tau}$ from the time series of acoustic pressure fluctuations ($p'$). Here, $N$ is the length of the time series, $m$ is the embedding dimension, and $\tau$ is the time delay for which the average mutual information of the signal is minimum \cite{fraser1986independent}. Then, we calculate the correlation sum  as follows:  
\begin{equation}
C(\epsilon) = \frac{2}{N(N-1)} \sum_{j=1}^{N} \sum_{k=j+1}^{N} \Theta(\epsilon - ||s_j - s_k||)
\end{equation}
where $\Theta$ is the Heaviside function defined as:\\
\vspace{5mm}
$\Theta(x)=\begin{cases}
1,& x>0, \\
0,& x\leq0,
\end{cases}$\\
and $\epsilon$ is the threshold distance. The sum counts the number of pairs $(s_j, s_k)$, whose distance is smaller than $\epsilon$. In the limit of infinite data $(N \to \infty)$ and $\epsilon \to 0$, $C(\epsilon)$ scales as a power law i.e., $C(\epsilon) \propto \epsilon^D$. We can define the correlation dimension $(D)$ from the correlation sum as follows:
\begin{equation}
    d(N, \epsilon) = \frac{\partial (\ln C(\epsilon, N))}{\partial (\ln(\epsilon))},
\end{equation}
\begin{equation}
    D = \lim_{\hspace*{2mm} N\to\infty \hspace*{2mm} \epsilon\to0} d(N,\epsilon).
\end{equation}
The value of $D$ can be used to estimate the dynamical state of a system. In general, the value of $D$ is noninteger for the strange attractor, and it is an integer and approximately one for periodic attractors \cite{mcmahon2017insights}.\\
It is sufficient to choose a value of $m$ greater than the box dimension to estimate the value of $D$ \citep{doi:10.1142/S0218127493000647}. To ensure the convergence of $c(\epsilon)$, we compute it for embedding dimensions ranging from $5-7$.  In figure~\ref{corr_sum}, we plot the variation of $C(\epsilon)$ with increasing $\epsilon$ at three embedding dimensions in a logarithmic plot for the states I to V. We apply a power law fit to the straight-line portion to obtain the correlation dimension ($D$).\\
\hspace*{10mm}For state I, the value of $D$ is $4.13$ (Fig.~\ref{corr_sum}\textcolor{blue}{a}). This high fractional value of $D$ indicates the necessity of a high-dimensional structure to describe the underlying chaotic dynamics. For the state II, the value of $D$ is $3.87$ (Fig.~\ref{corr_sum}\textcolor{blue}{b}), reflecting the need for a high-dimensional structure to capture the dynamics of epochs of high-amplitude periodic oscillations amidst epochs of low-amplitude aperiodic fluctuations (Fig.~\ref{FFT_PS}\textcolor{blue}{b}). The dynamics in the phase space indicate that the system switches between low-amplitude aperiodic fluctuations and high-amplitude periodic oscillations (Fig.~\ref{FFT_PS}\textcolor{blue}{l}), characteristics of the state of intermittency.\\
\hspace*{10mm}For state III, the reconstructed phase space exhibits complex folding and stretching (Fig.~\ref{FFT_PS}\textcolor{blue}{m}), requiring a high value of dimension to fully describe the dynamics. The observed value of $D$ of $2.89$ (Fig.~\ref{corr_sum}\textcolor{blue}{c}) suggests a complex structure, although less intricate than chaos and intermittency.\\
\hspace*{10mm}For state IV, the reconstructed phase space is found to be a thick ring (Fig.~\ref{FFT_PS}\textcolor{blue}{n}). The trajectory during a few acoustic cycles deviates from the mean, resulting in a large thickness. This is reflected by a value of $D$ of $1.62$ (Fig.~\ref{corr_sum}\textcolor{blue}{d}), indicating the need for an additional effective dimension to capture the noisy behavior. These noisy periodic oscillations with amplitude modulations are also called noisy limit cycle oscillations.\\
\hspace*{10mm}For state V, the reconstructed phase space exhibits a thin ring structure (Fig.~\ref{FFT_PS}\textcolor{blue}{o}), indicating periodic oscillations with fewer amplitude modulations. The value of $D$ is observed to be $1.08$ (Fig.~\ref{corr_sum}\textcolor{blue}{e}), which is very close to the integer value $1$. These clean periodic oscillations essentially clean limit cycle oscillations.\\
\hspace*{10mm}The results of $0-1$ test show that state III is nonchaotic and nonperiodic. Additionally, the value of $D$ falls between those of the chaotic and periodic states suggesting fractal characteristics. These findings imply that state III may be a strange nonchaotic attractor (SNA), which is a nonchaotic fractal. To confirm this, we perform a singular continuous spectrum analysis.
\paragraph{Singular continuous spectrum}
 In dynamical systems, the power spectrum is continuous during chaotic dynamics and discrete during regular dynamics. A continuous spectrum displays a broadband distribution, while sharp peaks characterize a discrete spectrum. The singular continuous spectrum is a mixture of both continuous and discrete spectrums. For the time series of acoustic pressure fluctuations $p'$, the singular continuous spectrum is defined as \citep{Pikovsky_1994, PhysRevE.56.1623},
\begin{equation}
    X(\alpha, N) = \sum_{k=1}^N p'(k) e^{2 \pi i k \alpha}.
    \label{SCSA}
\end{equation}
Here, $\alpha = \frac{\sqrt{5}-1}{2}$, the golden mean ratio, and $N$ is the length of the time series of $p'$. $X(\alpha, N)$ is a complex value. For chaotic signals, the power of the signal $|X(\alpha, N)|^2$ is proportional to $N$, indicating a continuous (broadband) spectrum. For periodic signals, the power of signal $|X(\alpha, N)|^2$ is proportional to $N^2$, illustrating a discrete (sharp) spectrum. For the state of SNA, the spectrum exhibits a singular continuous spectrum. For SNA the power of signal $|X(\alpha,N)|^2$ is proportional to $N^\gamma$ and the value of $\gamma$ ranges between $1$ and $2$.\\
\begin{figure}
\includegraphics[scale =0.5]{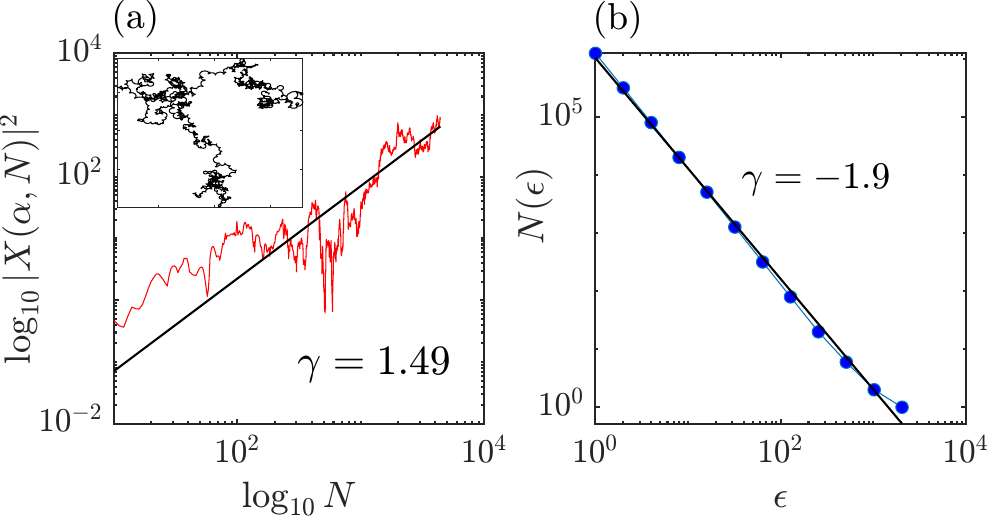}
\caption{\label{SCS} Singular continuous spectrum analysis of acoustic pressure oscillations ($p'$) for state III. (a) $\log_{10}|X(\alpha,N)|^2$ vs $\log_{10}N$. We have $|X(\alpha, N)|^2$ $\sim$ $N^{1.49}$. The subplot shows the corresponding trajectory (Re$X$, Im$X$), which is a fractal. (b) fractal dimension of $1.9$ calculated using box-counting method following a power law with increasing box size. }
\end{figure} 
\hspace*{10mm}We apply singular continuous spectrum analysis for the state III. We observe that the value of $\gamma$ is $1.49$ (Fig.~\ref{SCS}\textcolor{blue}{a}) and the trajectory of $X(\alpha, N)$ on the complex plane is a fractal as shown in the subplot of figure~\ref{SCS}\textcolor{blue}{a}. To confirm the fractal structure of the complex plane, we apply the box-counting dimension and calculate the fractal dimension of the structure.   
\paragraph{Box-counting dimension}
The concept of dimension is fundamental to fractal geometry, with self-similarity being a key characteristic of fractals \cite{mandelbrot1982fractal}. There are various definitions of fractal dimension (FD) \cite{doi:https://doi.org/10.1002/0470013850.ch3}, each suitable for specific contexts. Among the methods available to estimate FD, the box-counting method is frequently used due to its simplicity \cite{peitgen2004chaos}. The FD can be calculated using the method as follows:
\begin{equation} \label{bc}
    FD = - \lim_{\epsilon \to 0} \frac{\log(N(\epsilon))}{\log(\epsilon)},
\end{equation}
where $N(\epsilon)$ is the number of boxes of size $\epsilon$ required to completely cover the structure. The FD is estimated from the least square linear fit of log-log plot of $N(\epsilon)$ against $\epsilon$.\\
\hspace*{5mm} We apply the box-counting method on the image of the structure in the complex plane shown in the subplot of figure~\ref{SCS}\textcolor{blue}{b} and observe that the FD of the image is $1.9$ (Fig.~\ref{SCS}\textcolor{blue}{c}). This fractional value of FD indicates that the trajectory (Re$X$, Im$X$) in the complex plane is fractal in nature, confirming the presence of SNA.\\
\hspace*{5mm} In summary, we report the state of SNA in a self-organized system comprising turbulent reactive flow. We observe the state of SNA prior to the emergence of periodic oscillations from chaos. We distinguish the dynamical states present during the transition using tools from nonlinear dynamics, such as phase space reconstruction, $0-1$ test, and correlation dimension test. We substantiate the existence of the state of SNA with singular continuous spectrum analysis. The presence of SNAs offers the advantage of robustness to noise over chaos in synchronization phenomena owing to their insensitivity to initial conditions. This advantage of SNA and the possibility of SNAs in self-organized complex systems opens a new avenue for research. Further, exploration of spatiotemporal dynamics during the state of SNA could provide deeper insights into synchronization and other nonlinear phenomena.\\

We thank S. Anand, S. Thilagaraj, G. Sudha, M. Raghunathan, S. Sudarsanan, Ramesh S Bhavi, and Anaswara Bhaskaran for their help during the experiments. T.B. acknowledges the research assistantship from the Ministry of Human Resource Development, India, and the Indian Institute of Technology Madras. S.T. acknowledges the support from Prime Minister's Research Fellowship, Govt. of India. R.I.S. acknowledges the funding from the Science and Engineering Research Board (SERB) of the Department of Science and Technology (DST) through a J. C. Bose Fellowship (No. JCB$/2018/000034/$SSC) and from the IOE initiative (No. SP$22231222$CPETWOCTSHOC).

\appendix

\nocite{*}

\bibliography{apssamp}

\end{document}